\documentclass[twocolumn,twoside]{svmultivs_gm} 
\usepackage{graphicx}
\usepackage{amsmath}
\title*{Extracting Geodetic Data from GNSS-VLBI Co-Observation}
\subtitle{A Preliminary Local Tie Vector from Satellite Phase Delays}
\titlerunning{Geodetic Data from GNSS-VLBI}
\author{J.~Skeens$^1$, J.~York$^1$, L. Petrov$^2$, K.~Herrity$^1$, R. Ji-Cathriner$^1$, S. Bettadpur$^3$}
\authorrunning{Skeens et al.} 
\institute{1. Applied Research Laboratories, The University of Texas at Austin \\ 
           2. NASA Goddard Space Flight Center \\
           3. Center for Space Research, The University of Texas at Austin}
\ContactAuthorName{Joe Skeens}
\ContactAuthorTelephone{(614) 507-8198}
\ContactAuthorEmail{jskeens1@utexas.edu}
\NumberofInstitutions{3}
\InstitutionPostAddress{1}{10000 Burnet Rd, Austin, TX 78758}
\InstitutionCountry{1}{United States}
\InstitutionWebPage{1}{https://www.arlut.utexas.edu/}
\InstitutionPostAddress{2}{8800 Greenbelt Rd, Greenbelt, MD 20770}
\InstitutionCountry{2}{United States}
\InstitutionWebPage{2}{https://www.nasa.gov/goddard/}
\InstitutionPostAddress{3}{3925 W Braker Ln, Austin, TX 78759}
\InstitutionCountry{3}{United States}
\InstitutionWebPage{3}{https://www.csr.utexas.edu/}
\begin{document}  
\maketitle       
\keywords{GNSS, Local Ties, Near Field, Reference Frame, Phase Delay}
%
%
%
\section{Introduction}
Traditionally, local tie vectors have been determined by optical tie vector measurement, which uses a total station to range to 
reflectors mounted on space geodetic instruments. The total station measurements relate the positions of the optical reflectors and not
the reference points of geodetic measurements and thus require an additional model or transformation to the reference points
of the space geodetic instruments to realize the local tie vector. The accuracy of this transformation is difficult to verify.

We instead aim to produce local ties directly through geodetic measurements, whether through interferometric VLBI-style processing, or 
by differencing pseudorange and carrier phase measurements made through correlation of an incoming signal from a Global Navigation Satellite Systems (GNSS) 
satellite with a local replica generated from the satellite's unique pseudorandom noise (PRN) code.
With these GNSS antenna to VLBI radio telescope processing techniques, we can map geodetic measurements made through the radio frequency hardware
to the geodetic reference points via the same models and analysis techniques used in routine GNSS and VLBI processing. 
The ability to directly compare GNSS and VLBI analysis techniques also raises the possibility of
identifying systematic biases introduced in one technique and not the other. Additionally, because the GNSS antenna can capture observables from all
visible satellites simultaneously, we can leverage precise point positioning (PPP) or precise differential positioning (PDP) with other GNSS antennas to produce 
high quality a priori positions and clock biases that can be utilized when estimating a local tie vector.

We have previously demonstrated through VLBI-style processing that we can detect both GNSS satellites of multiple systems and 16 radio galaxies including compact active galactic nuclei \cite{skeens_23}.
More detail about the hardware and software of the High Rate Tracking Receiver as well as additional details on the configuration of our experiments and observing strategy can be found in that paper.

\section{Data}
\begin{figure}[htb!]         
  \includegraphics[width=\columnwidth]{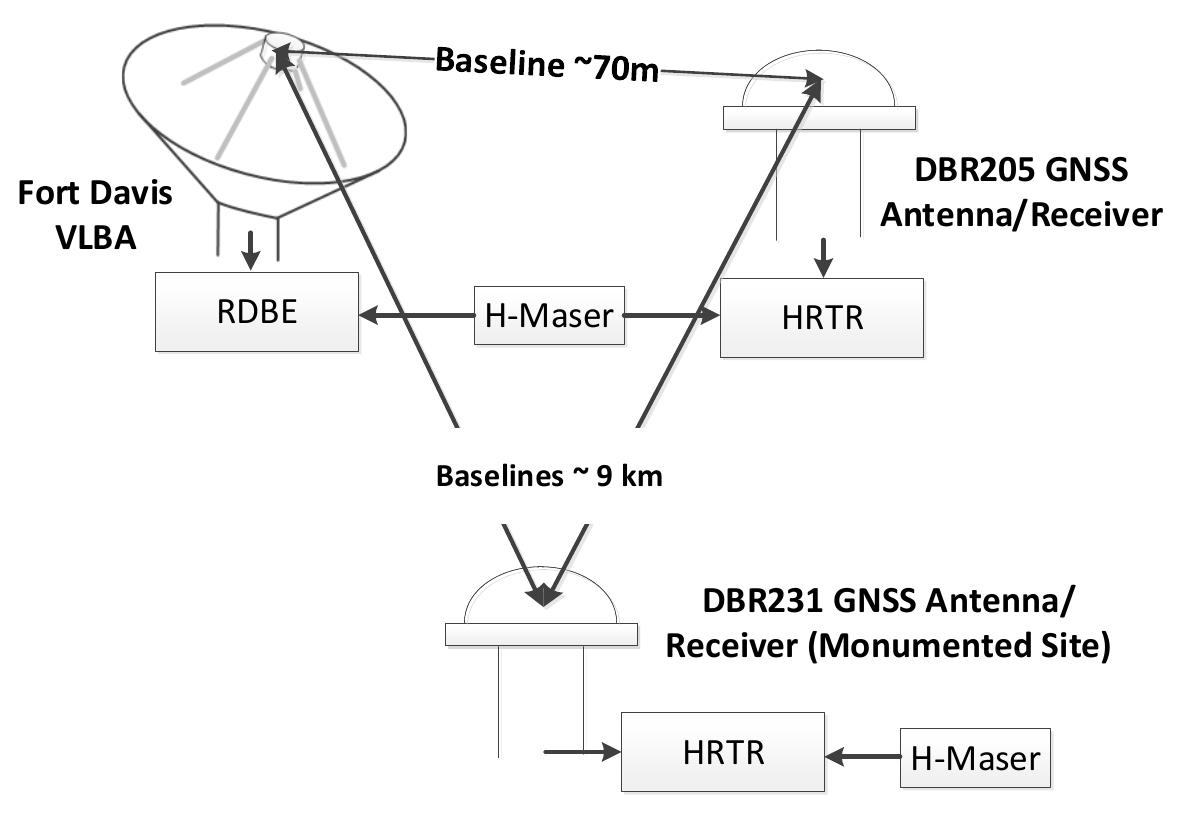}
  \caption{The setup of the data collection on January 25, 2023. One GNSS antenna was deployed at Ft. Davis, TX about 70~m from the VLBA telescope with a common clock feed.
    Another was deployed about 9~km away on a deep-drilled braced monument with a feed from the VGOS H-maser clock.}
  \label{fig:exp_setup}             
\end{figure}

%
Figure \ref{fig:exp_setup} shows the setup of a data collection we carried out on January 25, 2023 at the Very Long Baseline Array (VLBA) radio
telescope in Fort Davis, Texas. We deployed two Topcon CR-G5 GNSS antennas with High Rate Tracking Receivers (HRTRs), one at the Fort Davis site about 70~m from the
VLBA telescope on a tripod mount, and another at the University of Texas McDonald Geodetic Observatory (MGO) about 9~km away. Both GNSS antennas received feeds from 
hydrogen maser (H-maser) frequency standards. The antenna and receiver at Fort Davis, hereafter called DBR205, received a feed from the same clock used by the FD-VLBA radio telescope.
The GNSS antenna and receiver at MGO, hereafter called DBR231, received a feed from the H-maser at the VGOS station MACG012M and was deployed on a deep-drilled braced monument (DDBM). 

In total, we observed 19 GNSS satellites in 84 scans of about 30 seconds each taking place over about 4 hours. The satellites belonged to the GPS, Galileo, and BeiDou satellite systems. We scheduled the VLBA observations in sur\_sked, utilizing ephemerides produced from forward-propagated two-line element orbits with the SGP4 propagator. 
The radio telescope tracked only sidereal motion, thus the GNSS satellite was allowed to drift through the main beam of the telescope during each observation. 

\begin{figure}[htb!]
\includegraphics[width=1.05\columnwidth]{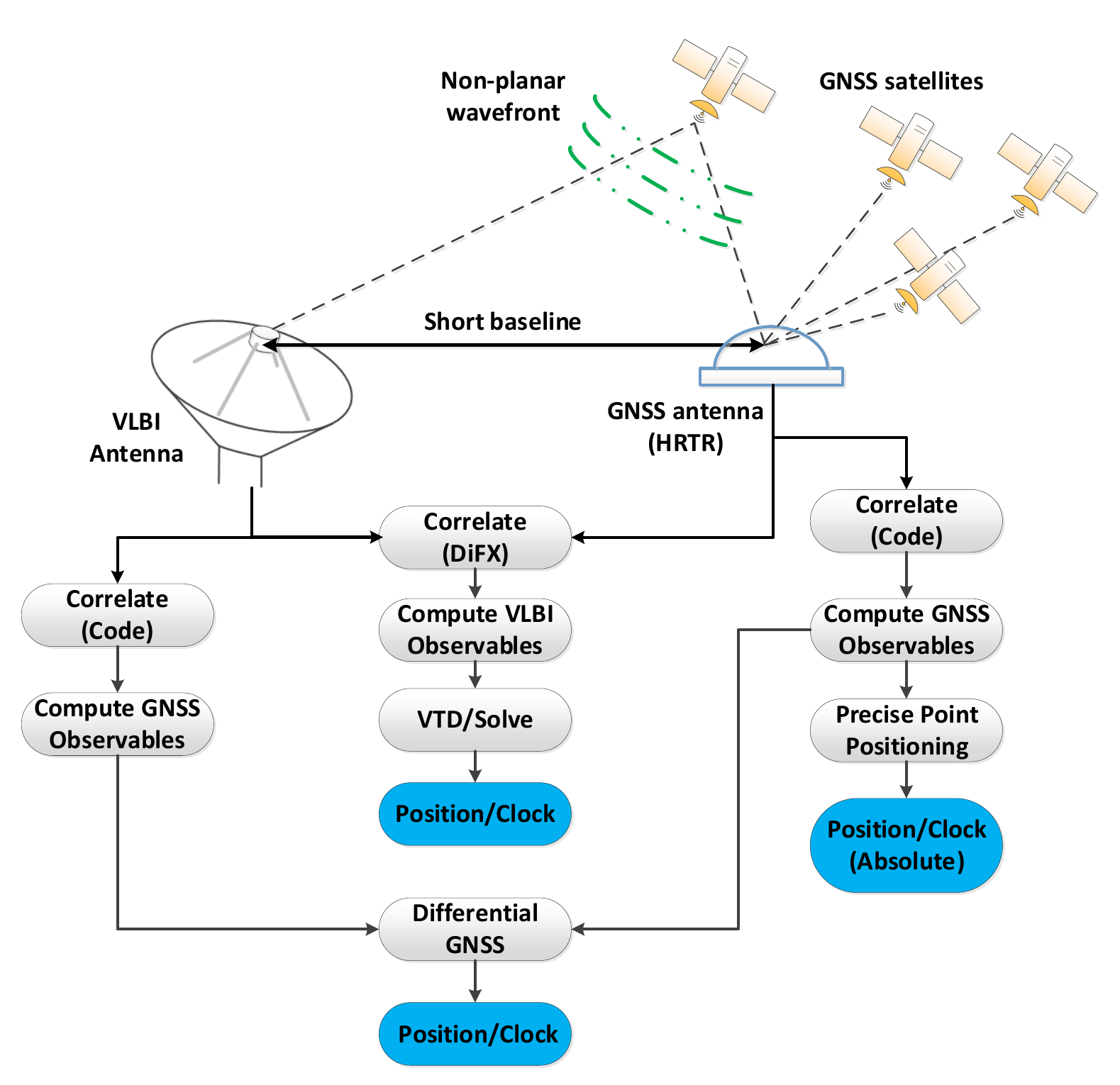}
\caption{The processing setup for geodetic observables collected on GNSS satellites. Correlation of the GNSS satellite signal against a local replica of the code allows for GNSS processing,
    while correlation against the baseband samples from another antenna allows for VLBI processing. The patch-style GNSS antenna allows for collection of GNSS observables for all visible satellites, 
    which can be used to produce an absolute position and clock solution. }
\label{fig:gnss_processing}
\end{figure}

Figure \ref{fig:gnss_processing} shows our processing strategy for observations of GNSS satellites. The VLBA radio telescope observes one satellite in a 30-second scan, 
while the two patch GNSS antennas record continuous baseband data containing the signals of all GNSS satellites above the horizon. 
The baseband samples from the GNSS antennas are correlated against a local replica of the 
PRN of each visible satellite to get pseudorange and carrier phase observables. Using these observables, we can perform a precise point positioning (PPP) or 
precise differential positioning (PDP) solution to get a high accuracy estimate of the position and clock of each of the GNSS antennas in the International Terrestrial Reference Frame (ITRF) as realized
by the International GNSS Service (IGS) in final ephemerides derived from a large network solution. We use the multi-GNSS ephemerides produced by the Astronomical Institute of the University of Bern's 
Center for Orbit Determination (COD). 

In the central branch of the figure, VLBI-style processing is carried out to estimate a differential position and clock function that represents the local tie vector between the GNSS antennas 
and the VLBI radio telescope. The stages of this analysis are, software correlation in DiFX \cite{difx} using a near-field interferometric model, fringe fitting in PIMA to
produce group and phase delays, and parameter estimation in a fork of Solve to estimate the local tie vector by holding the GNSS antenna position fixed and allowing the VLBA telescope position to vary.
At each stage of this VLBI analyis pipeline, the delay modeling is done with the analytical near-field delay model detailed in \cite{jaron_19} from satellite positions interpolated with basis splines
from the COD precise ephemerides.

The baseband samples from FD-VLBA can also be correlated against a local replica of the GNSS satellite that is currently being observed in the main beam of the radio telescope to produce pseudorange and carrier phase observables for that single satellite. 
However, it is important to note that this produces GNSS observables for only a single satellite per epoch, and thus cannot be used in traditional PPP or PDP to 
derive a position estimate. The observables can be differenced with the corresponding pseudorange and carrier phase for that satellite recorded by the GNSS antenna, 
meaning that a differential GNSS solution can be estimated in tandem with the VLBI processing. We have successfully produced pseudorange and carrier phase measurements from the FD-VLBA telescope for the
first time, and we are currently developing software to carry out this unique GNSS-style analysis. The differential positioning solution produced through this GNSS-style analysis should match that produced through VLBI analysis, and discrepancies may reveal systematic biases in either technique.

\section{Analysis}
Using the GNSS data captured before, during, and after co-observing with the FD-VLBA telescope, we can derive a precise point positioning solution for the two GNSS antennas/receivers, 
DBR205 and DBR 231 (Table \ref{tab:ppp_res}). In addition to these receivers, there are five contributing IGS GNSS antennas/receivers at MGO--MDO1, MGO2, MGO3, MGO4, and MGO5. The positions of these 
antennas derived through the same PPP processing for 24 hours of data are shown in the table. 
MDO1 and MGO2 are separated by about 40~m, while DBR231 and MGO4 are separated from MGO5 by about 70~m. The two clusters are separated by about 800~m, and MGO3 is about 450~m from the former cluster and 800~m from the latter.
The positions of the five IGS receivers have formal uncertainties on the order of 1~mm, which is likely an underestimation of the true position uncertainty. 
With only 4.5 hours of data, the position of DBR205 is not well-constrained through PPP alone.
DBR231 has 24 hours of data with pseudorange and carrier phase measurements recorded by hardware processing on the HRTR's Field Programmable Gate Array (FPGA). These observables
for DBR205 and DBR231 were only recorded for GPS L1 and L2 civil signals. The IGS receivers record observables for GPS, Galileo, and GLONASS, however, only GPS and Galileo were used in the PPP solutions.

\begin{table*}[htb!]
    \caption{Results of Precise Point Positioning for the International GNSS Service receivers MDO1 MGO2, MGO3, MGO4, MGO5, and the two deployed High Rate Tracking Receivers, DBR205 and DBR231.}
\begin{tabular}{|l|c|c|c|c|c|c|c|} \hline
    Receiver & Data length (hr) & X (m)& Y (m)& Z (m)& $\sigma_X$ (mm) & $\sigma_Y$ (mm) & $\sigma_Z$ (mm) \\
\hline
    DBR205     & 4.5& -1324070.4796 & -5332176.0169 & 3231921.7803 & 11.9& 11.5&  8.9\\ \hline
    DBR231     & 24 & -1330748.6520 & -5328115.2850 & 3236419.9321 &  2.0&  3.2&  1.8\\ \hline
    MDO1 (IGS) & 24 & -1329998.9947 & -5328393.3463 & 3236504.0576 &  1.2&  1.9&  1.1\\ \hline
    MGO2 (IGS) & 24 & -1329992.4399 & -5328416.6674 & 3236473.3342 &  0.5&  0.9&  0.5\\ \hline
    MGO3 (IGS) & 24 & -1330165.5451 & -5328541.9492 & 3236118.3457 &  0.6&  1.0&  0.6\\ \hline
    MGO4 (IGS) & 24 & -1330807.0959 & -5328078.7274 & 3236452.4333 &  0.6&  1.0&  0.6\\ \hline
    MGO5 (IGS) & 24 & -1330810.8347 & -5328115.5555 & 3236393.6386 &  0.6&  0.9&  0.6\\
\hline
\end{tabular}
\label{tab:ppp_res}
\end{table*}

The addition of five IGS receivers improves the robustness of the precise differential positioning solution that we used to determine the position of the deployed GNSS antenna/receiver DBR231 (Table \ref{tab:pdp_res}). 
In this differential positioning analysis, the position of the antenna MDO1, determined by PPP to a formal uncertainty of about 2~mm with 24 hours of data, is used as the reference to estimate the positions of the remaining five GNSS antennas.
DBR205 is not included in this differential solution because it would limit the data to 4.5 hours rather than 24 hours.
The estimated positions of all included antennas have formal uncertainties below 1~mm in their Cartesian position components.

\begin{table*}[htb!]
\caption{Results of Precise Differential Positioning for the deployed High Rate Tracking Receiver DBR231 and other McDonald Geodetic Observatory IGS receivers referenced to the IGS receiver MDO1.}
\begin{tabular}{|l|c|c|c|c|c|c|c|} \hline
    Receiver &  X (m)& Y (m)& Z (m)& $\sigma_X$ (mm) & $\sigma_Y$ (mm) & $\sigma_Z$ (mm) \\
\hline
    DBR231 & -1330748.6540 & -5328115.2871 & 3236419.9350 & 0.5& 0.5& 0.3\\ \hline
    MGO2   & -1329992.4448 & -5328416.6691 & 3236473.3350 & 0.5& 0.6& 0.4\\ \hline
    MGO3   & -1330165.5517 & -5328541.9490 & 3236118.3473 & 0.5& 0.6& 0.4\\ \hline
    MGO4   & -1330807.0979 & -5328078.7290 & 3236452.4336 & 0.5& 0.6& 0.4\\ \hline
    MGO5   & -1330810.8380 & -5328115.5571 & 3236393.6400 & 0.5& 0.6& 0.4\\ 
\hline
\end{tabular}
\label{tab:pdp_res}
\end{table*}

The PDP/PPP positions of DBR231/DBR205 are used as the a priori positions in a VLBI network analysis including DBR205, DBR231, and FD-VLBA in which DBR231 is held as the reference. 
We expect that there will be some discrepancy between the estimated position of the VLBA antenna 
in the IGS-distributed International Terrestrial Reference Frame (ITRF) and the a priori FD-VLBA position in the VLBI reference frame.
Because there are three baselines, we resolve all phase delay ambiguities simultaneously, and we have access to phase delay closure measurements to assess the quality of this geodetic analysis.

\begin{figure}[htb!]         
  \includegraphics[width=1.05\columnwidth]{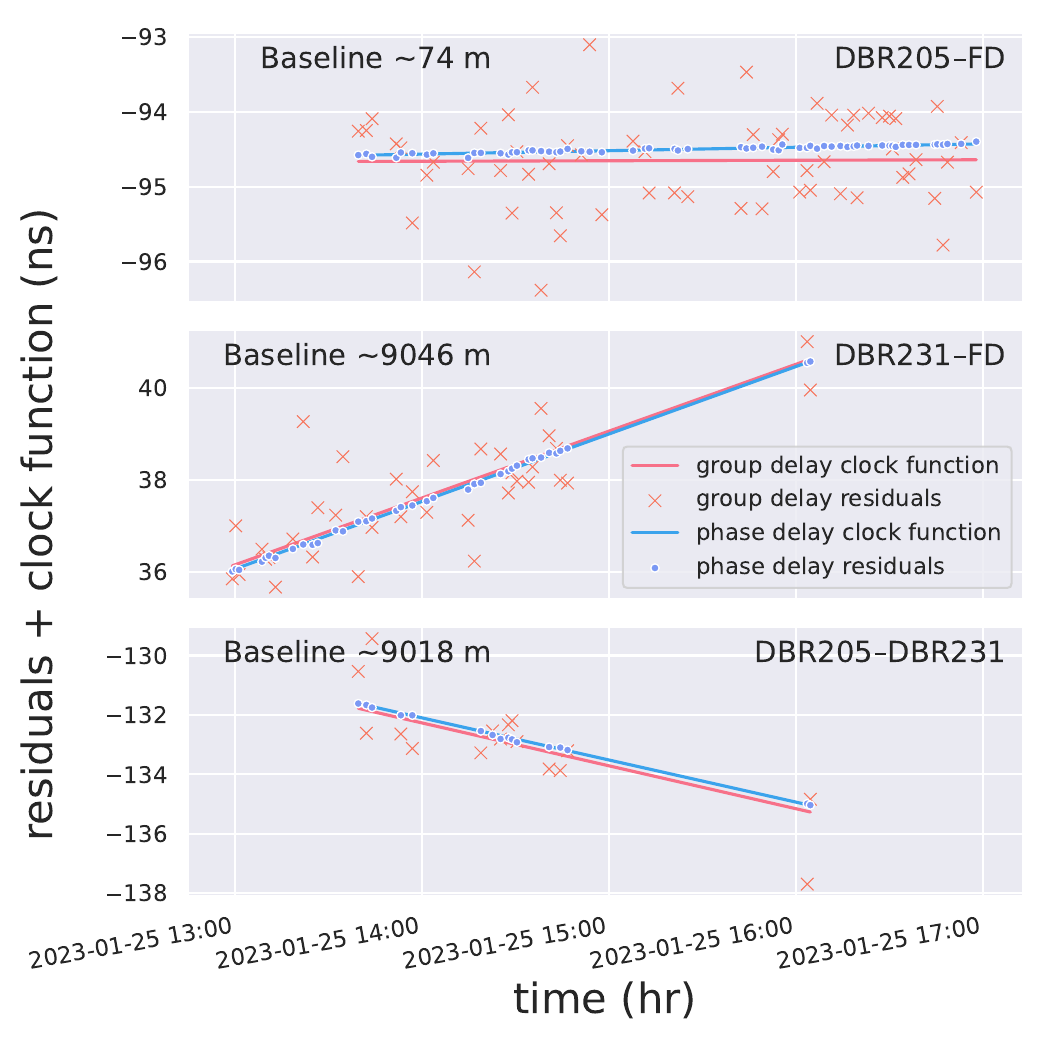}
    \caption{The postfit residuals plus clock function for each of the three VLBI baselines: DBR205--FD-VLBA (top), DBR231--FD-VLBA (middle), DBR205-DBR231 (bottom). The WRMS for the group delay residuals is 614~ps, 766~ps, and 1046~ps. The WRMS for the phase delay residuals is 18~ps, 49~ps, and 44~ps respectively with an ambiguity interval of 650~ps. }
  \label{fig:postfit_resids}             
\end{figure}

Figure \ref{fig:postfit_resids} shows the group and phase delay residuals on each of the three baselines overlaid on the estimated differential clock function. 
Due to some software and hardware complications that we expect to resolve in future data collections, the DBR205 baseband data did not begin until about a half hour into the VLBA observation, and the DBR231 baseband data ended about an hour before the end of the scheduled VLBA observation collection time with a short resumed segment of baseband data at about 16:00 UTC. 
We excluded observations that did not pass a quality control test ensuring that there is no visible slope in the plot of fringe phase against frequency or time off-source as evidenced by fast varying, low amplitude phase for a segment of the fringe time series. 
After this procedure, no outliers were identified or removed amongst the remaining group and phase delay measurements.

Verification of phase delay ambiguity resolution is a major challenge for this effort, as GNSS signals have relatively narrow bandwidths, from 2-4~MHz for civil signals to 15-20~MHz for the fastest chipping military signals. Fortunately,
in VLBI-style processing, treating the signals as random noise allows us to use the wider bandwidth signals without knowing the secret chipping sequence of restricted signals. Even still,
the weighted root-mean-square (WRMS) group delay residuals are on the order of 600-700~ps, about the same as the 635~ps phase ambiguity interval of the 1575~MHz carrier frequency. 
However, after simultaneous ambiguity resolution, the WRMS of the phase delay residuals is on the order of 20-50~ps, less than a tenth of the ambiguity interval. 
Ambiguity resolution was made significantly simpler by the short baseline and common clock of the stations DBR205 and FD-VLBA.
This allowed us to estimate only a single linear clock function and no atmospheric parameters, vastly reducing the kind of overfitting that can result in incorrect phase delay ambiguity resolution. After phase delay resolution, the RMS phase delay closure for scans common to all three baselines is 0.05~ps. 

Table \ref{tab:vlbi_pos} shows the positions estimated in the network VLBI solution along with associated formal uncertainties. 
We also ran a second dual-frequency PDP solution between DBR205 and DBR231 using the DBR231 PDP position as the reference.
Here, the DBR205--DBR231 baseline can be used to compare the PDP solution directly to the VLBI solution. 
The dual-frequency PDP resolved the ambiguities using a separate double-differenced scheme, so the agreement between these two positions to well within a wavelength (about 19~cm) lends credence to our having correctly resolved the phase delay ambiguity.

\begin{table*}[htb!]
\caption{The VLBI differential positioning solution using DBR231 as the reference position and estimating a single linear clock interval compared to PDP and VLBI a priori.}
\begin{tabular}{|l|c|c|c|c|c|c|c|c|c|} \hline
    Receiver &  X (m)& Y (m)& Z (m)& $\sigma_X$ (mm) & $\sigma_Y$ (mm) & $\sigma_Z$ (mm) & $\sigma_\text{clock}$ (ns) &  $\sigma_\text{rate}$ (sec/sec)  \\
\hline
    DBR205 (group delay)  & -1324070.5140 & -5332175.7838 & 3231921.7340 &  76.5&   192.0&  102.0 & 0.465  & 4.922$\cdot10^{-14}$\\ \hline
    DBR205 (phase delay)  & -1324070.4674 & -5332176.0298 & 3231921.7960 &  3.9&   10.2&  5.4 & 0.023  & 0.251$\cdot10^{-14}$\\ \hline
    DBR205 (PDP)          & -1324070.4781 & -5332176.0011 & 3231921.7985 &  3.4&    4.6&  4.1 & --     & --                  \\ \hline
    FD-VLBA (group delay) & -1324009.4957 & -5332181.8279 & 3231962.2923 &  67.0&  180.8& 92.5 & 0.408  & 4.500$\cdot10^{-14}$\\ \hline
    FD-VLBA (phase delay) & -1324009.4486 & -5332181.9826 & 3231962.3565 &  3.8&    10.1& 5.2 & 0.023  & 0.246$\cdot10^{-14}$\\ \hline
    FD-VLBA (a priori)    & -1324009.4541 & -5332181.9548 & 3231962.3693 &   --&     --&   -- & --     & --                   \\
\hline
\end{tabular}
\label{tab:vlbi_pos}
\end{table*}

\begin{figure}[htb!]         
  \includegraphics[width=\columnwidth]{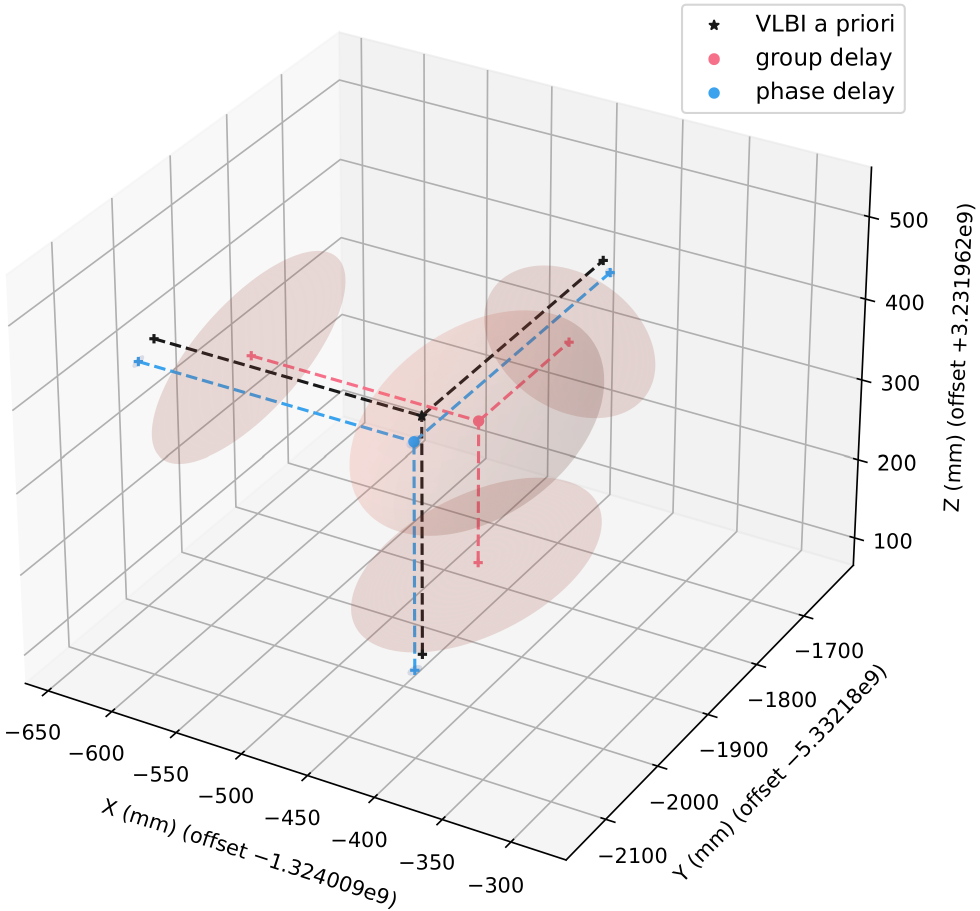}
    \caption{The estimated position of the FD-VLBA radio telescope using group delays (red) and phase delays (blue) with DBR231 position fixed compared to an a priori estimated position (black) from a Radio Fundamental Catalogue (RFC) solution.}
  \label{fig:pos_vlba_231}             
\end{figure}

Figure \ref{fig:pos_vlba_231} shows a graphical representation of the estimated positions and ellipsoidal uncertainties for FD-VLBA with projections to the X-Y, X-Z, and Y-Z planes. 
As expected, the FD-VLBA position estimated in the IGS-distributed reference frame has a cm-level difference with the VLBI a priori. 
This indicates that even this preliminary local tie vector has useful information. 

 
\section{Conclusions}
In this paper, we have demonstrated a preliminary local tie vector between deployed GNSS antennas/receivers and a VLBI radio telescope. We have significant evidence that we have correctly resolved the phase delay ambiguity, and formal uncertainties suggest that we have constrained the position of the FD-VLBA radio telescope to about 1~cm. 
At this stage, we are unable to evaluate the reliability of the solution, and the uncertainty does not account for any systematic errors that we will investigate in the future.
However, these initial results are extremely encouraging and suggest that this technique may have many useful applications.
Planned future experiments will focus on improvements in two areas:
first, we will collect longer time series of geodetic observables and use monumented GNSS antenna locations to reduce uncertainties.
Second, we will analyze the repeatability of the local tie vectors to obtain a more reliable method of assessing local tie accuracy. 

Another concern that has been raised in the community is that these local tie vectors are necessarily through the L band feed of participating VLBI antennas, which is not the typical S/X band feed used in geodetic VLBI experiments. 
To first order, this is not a problem for the technique, as the difference in phase center location does not depend on antenna orientation.
The contribution of this effect is thus not distinguishable from clock error and is absorbed in the estimate of the clock function during data analysis.
However, if small discrepancies exist that affect the estimated position, a geodetic VLBI experiment in L band could be conducted close in time
to a traditional S/X experiment such as an intensive to investigate its magnitude.

\end{document}